\documentstyle[aps,multicol,pra,epsfig]{revtex}
\def\bea{\begin{eqnarray}}
\def\eea{\end{eqnarray}}
\def\be{\begin{equation}}
\def\ee{\end{equation}}
\begin{document}
\draft
\author{A. Trombettoni$^{1,2}$, A. Smerzi$^{1,2}$ and A.R. Bishop$^1$}
\address{
 $^1$ Theoretical Division and Center for Nonlinear Studies,
        Los Alamos National Laboratory,
        Los Alamos, NM 87545 USA\\
$^2$ Istituto Nazionale di Fisica per la Materia and
International School for Advanced Studies,\\
 via Beirut 2/4, I-34014, Trieste, Italy}
\title{Superfluidity versus Disorder in the Discrete Nonlinear
Schr\"odinger Equation}
\date{\today}
\maketitle
\begin{abstract}
We study the discrete nonlinear
Schr\"odinger equation (DNLS) in an annular geometry with on-site defects.
The dynamics of a traveling plane-wave maps onto 
an effective ''non-rigid pendulum'' Hamiltonian.
The different regimes include the complete reflection and 
refocusing of the initial wave, 
solitonic structures, and a
superfluid state. 
In the superfluid regime, which occurs above a critical value of nonlinearity,
a plane-wave travels coherently through the 
randomly distributed defects. This superfluidity criterion for the DNLS
is analogous to (yet very different from)  
the Landau superfluidity criteria in translationally 
invariant systems.
Experimental implications
for the physics of Bose-Einstein condensate gases trapped in optical
potentials and of arrays of optical fibers are discussed.
\end{abstract}
\begin{multicols}{2}

Studying the interplay between nonlinearity and disorder
has become a fundamental
issue of the last decades in the study of many
physical and biological systems 
(both discrete and continuous) \cite{scott99}.
It is well known that nonlinearity $or$ disorder
may lead to localized excitations -
'solitonic' structures \cite{sanchez88} 
and 'Anderson localization' \cite{ramakrishnan86}, respectively.
However, the dynamical properties of the system,
when both nonlinearity $and$ disorder are present, 
are still challenging theoretical investigations. 
This problem is, moreover, of central experimental relevance, 
since impurities can be
reduced but never completely eliminated.
In particular, it can be asked if
random defects will (and, if so, how) destroy
the propagation of traveling plane waves or localized excitations
(allowed by the nonlinearity), and what are the conditions for crossing
from a "superfluid" regime with propagation (and coherence)
preserved (due to, for instance, a large nonlinearity), to a "normal" regime
with disorder induced damping.

Here, we consider the dynamical properties of the DNLS
in an annular geometry and
in the presence of impurities. We
choose the DNLS for two main reasons: 1) it has all the required ingredients:
nonlinearity, disorder and discreteness; 
2) real physical systems, such as optical fibers and Bose-Einstein
condensates in deep optical lattices, map onto the DNLS and provide an
ideal experimental framework.  
The annular geometry is paradigmatic for studying the nature of
superfluidity \cite{leggett01}, and
allows a clear comparison between the discrete and 
continuous (translationally invariant) limits.

With regard to optical fibers, a typical experimental setup is 
an array of one dimensional nonlinear coupled waveguides
\cite{eisenberg98}. As the light
propagates through the array, the coupling
induces an exchange of power among the single waveguides.
In the low power limit (i.e. when the nonlinearity is negligible),
the optical field spreads over the whole array.
Upon increasing the power, the output field narrows
until it is localized in a few waveguides, and discrete solitons can 
finally be observed \cite{eisenberg98,morandotti99-1}.
The evolution of $E_n(\tau)$, the electrical field in the
$n$th waveguide, as a function of the position, $\tau$,
is governed by the DNLS Eq.(\ref{DNLS}).
In this case $\Lambda$ is proportional to the Kerr nonlinearity
and the on-site potentials $\epsilon_n$ are
the effective refraction indices of the individual waveguides.
In \cite{morandotti99-2}  
a linearly growing $\epsilon_n$ was realized,
which allowed the observation of Bloch oscillations.

Another significant physical system described by the DNLS
is a Bose-Einstein condensate (BEC) gas confined in a deep
optical lattice. 
A one dimensional optical lattice can be created experimentally by
a far-detuned, retro-reflected laser beam 
\cite{anderson98,burger01,cataliotti01,morsch01}.
The condensate wavefunction $\psi(\vec{r},t)$ obeys the
Gross-Pitaevskii equation \cite{dalfovo99} which is a (continuous)
nonlinear Schr\"odinger equation, with the nonlinearity arising from
the interatomic interaction. When the heights of the
intra-well barriers of the periodic optical potential are much
higher than the
condensate chemical potential, 
the system can be mapped onto the DNLS Eq.(\ref{DNLS}) \cite{trombettoni01},
with $\psi_n$ the condensate amplitude in the $n$th well and
$\epsilon_n$ proportional to any external field superimposed on the lattice.
In \cite{anderson98} a coherent output of matter waves was created by a
vertical optical array (with the
gravity gradient providing $\epsilon_n
\propto n$). In
\cite{cataliotti01}, the magnetic trap and the laser beams were turned
on in a superimposed harmonic magnetic field
($\epsilon_n \propto n^2$), allowing
the direct observation of coherent (Josephson) condensate oscillations
governed by Eq. (\ref{DNLS}).

The DNLS is (in dimensionless units):
\begin{equation}
\label{DNLS}
i  \frac{\partial \psi_n}{\partial \tau} = - \frac{1}{2}
(\psi_{n-1}+\psi_{n+1}) + (\epsilon_n+ \Lambda \mid \psi_n \mid
^2)\psi_n,
\end{equation}
where $\Lambda$ is the nonlinear coefficient and $n=1,\cdots,N$ ($N$
number of sites).
In the physical systems we have discussed, the defects $\epsilon_n$
can be spatially localized or extended. For instance, 
the impurities in optical fibers  can
be induced by different (possibly random) effective refraction indexes
of the guides
or with varying spatial separations between them.
In BEC's the defects can be created
with additional lasers and/or magnetic fields; the presence of a thermal
component can also be phenomenologically modeled, in some limits,
by a random distribution of defects.

We consider, first, the DNLS with a single impurity
$\epsilon_n=\epsilon \,\delta_{n,\bar{n}}$
at the site $\bar{n}$, and 
study the propagation of a plane wave $\psi_n(\tau=0)=e^{i k n}$. 
In the following
we assume $\Lambda > 0$
(which corresponds to a repulsive interatomic interaction in BEC's,
as is the case for $^{87}Rb$ atoms). Note, however, that Eq.(\ref{DNLS})
is invariant with respect to the transformation
$\Lambda \to -\Lambda$, $\epsilon_n \to -\epsilon_n$, 
and $\psi_n \to \psi_n^{\ast} e^{i \pi n}$.
Since we consider periodic boundary conditions
(due to the annular geometry), 
we have $k=2\pi l/N$ with $l$ integer ($l=0,1,\cdots,N-1$).

In the translationally invariant limit of the DNLS,
given by the continuum nonlinear Schr\"odinger equation (CNLS), 
a well know argument suggested by
Landau implies that 
superfluidity occurs when the speed is smaller than
the sound velocity (for weak perturbations).
A simple derivation of the Landau critical velocity
in the CNLS was recently proposed in \cite{leggett01}
considering an annular geometry with a single (small) impurity.
The key point was to map the problem of the propagation of a plane wave 
to a Josephson-like Hamiltonian.
The superfluid regime is allowed by the nonlinearity, which provided 
an effective energy barrier against the creation of elementary
excitations with momentum $k+q,k-q$ (with $q$ arbitrarily small)
which would dissipate the energy of the 
incident wave having momentum $k$.

This scenario is completely changed by
discreteness.
First, it is well known that when $\cos{k}<0$ the system becomes
modulationally unstable \cite{kivshar92}.
Stability analysis reveals that the eigenfrequencies of
the linear modes become
imaginary driving an exponential growth
of small perturbations.
This modulation instability disappear, for $\Lambda > 0$,
in the CNLS limit.
Let us consider, then, the case in which $\cos k > 0$.
In the absence of the impurity, superpositions of 
rotational states with opposite wave-vectors 
$k, -k$ are degenerate. The defect
splits the degeneracy by coupling the two $k, -k$ waves,  
very much as the tunneling barrier does in a double well potential,
with "left" and "right" localized states.   
Therefore, the relative population of the two waves oscillates
according to an effective 
(generalized \cite{smerzi97}) Josephson Hamiltonian. 
These Josephson regimes are preserved as far as the splitting in energy
induced by the defect $\sim \epsilon$ 
is much smaller than the energy gap between different rotational states  
$\sim 2 \pi \sin{k}$. In this limit, we can write the 
wavefunction $\psi_n(\tau)$ as  
\begin{equation}
\psi_n(\tau) = A(\tau) e^{i k n} + B(\tau)e^{- i k n}.
\label{ansatz}
\end{equation}
In the following, we set $A,B=\sqrt{n_{A,B}(\tau)}
e^{i \phi_{A,B}(\tau)}$, $z=n_A-n_B$ and $\phi=\phi_A-\phi_B$.
We will compare the numerical solution of (\ref{DNLS})
with the analytical solution of
(\ref{z-phi}) obtained from
the ansatz (\ref{ansatz}).

The two-mode Eq.(\ref{ansatz}) can be extended 
to the case of a time-dependent, 
arbitrary (including random) distribution of defects, 
with $\epsilon$ replaced, as shown below, 
by an effective impurity strength.
Furthermore, when the initial wave function is given by the sum
of multiple waves 
$\psi_n(0)= \sum_j A_j e^{i k_j n}$,
the ansatz (\ref{ansatz}) can
be straightforwardly generalized
so long as the quasi-momentum 
distributions peaked around $k_j$
do not overlap. The collision of a soliton 
with a single impurity has been
studied, from a different perspective, 
in \cite{scharf91}. A numerical analysis of the propagation of 
plane waves across a segment with defects was made in \cite{molina00}.  

Let us now derive the equations of motion. 
We define an effective Lagrangian as
${\cal L}= \sum_n  i \dot{\psi}_n \psi_n^\ast - \cal{H}$,
where ${\cal{H}} = {\sum_n} [ -\frac{1}{2} ( \psi_n \psi^\ast_{n+1}
+ \psi^\ast_n \psi_{n+1} ) + \epsilon_n \mid
\psi_n\mid^2 + {\Lambda \over 2} \mid\psi_n\mid^4  ]$
(both the Hamiltonian ${\cal H}$ and the norm $\sum_n \mid \psi_n \mid
^2=N$ are conserved).
The Euler-Lagrange equations
$\frac{d}{d t} \frac{\partial {\cal L}}{\partial {\dot q}_i} =
\frac{\partial {\cal L}}{\partial  q_i} $ for the variational
parameters $q_i(\tau)=n_{A,B}, \phi_{A,B}$ give the following equations:
\begin{mathletters}
\label{z-phi}
\begin{eqnarray}
\dot{z}&=&-\frac{2 \epsilon}{N} \sqrt{1 - z^2} \sin{\phi} \\
\dot{\phi} &=& \frac{2 \epsilon}{N} \frac{z}{ \sqrt{1 - z^2}}
\cos{\phi} + \Lambda z,
\end{eqnarray}
\end{mathletters}
with the replacement $\phi + 2 k \bar{n} \to \phi$.
The total (conserved) energy is:
\begin{equation}
H = \frac{\Lambda z^2}{2} - \frac{2 \epsilon}{N}
\sqrt{1-z^2} \cos{\phi}
\label{ham}
\end{equation}
and the equations of motion (\ref{z-phi})
can be written in the Hamiltonian form
$\dot{z} = -\frac{\partial H}{\partial \phi}$ and
$\dot{\phi} = \frac{\partial H}{\partial z}$
with $z$ and $\phi$ canonically conjugate variables.

The Eqs.(\ref{z-phi}) have been studied in very different contexts,
including polaron dynamics, where the dimer Eqs.(\ref{z-phi})
has been solved analytically \cite{kenkre86}, and in 
the Josephson dynamics of two weakly coupled
Bose-Einstein condensates \cite{smerzi97}.
Eqs.(\ref{z-phi}) are those of a non-rigid pendulum:
$\phi$ is the angular position and $z$ its conjugate momentum.
The non-rigidity of the pendulum is due to its momentum dependent length.

The pendulum phase portrait, $z$-$\phi$, has been studied in
\cite{smerzi97}.  
Let us briefly recall the main results.
We have a) oscillations around $<\phi>=0$ and $<z>=0$ ({\em $0$-states});
b) oscillations around $<z> \neq 0$ with running
phase $<\phi> \propto t$ ({\em self-trapped states});
c) oscillations around $<z>=0$
and $<\phi>=\pi$ ({\em $\pi$-states});
d) oscillations about $<z> \neq 0$
and $<\phi>=\pi$ ({\em self-trapped $\pi$-states}).
Here $<\cdots>$ stands for a time average.
The stationary points of Eqs.(\ref{z-phi}), $z=0$, $\phi=0$ 
(i.e. $A=B$) and $z=0$, $\phi=\pi$ ($A=-B$), correspond 
to time-independent solutions of Eq.(\ref{DNLS}), 
$\psi_n=2 \cos{k n}$ and $\psi_n=2 i  \sin{k n}$, respectively.

To understand the meaning of these regimes in our system,
we observe that the angular momentum is
$L(\tau)=i\sum_n(\psi_n \psi_{n+1}^{\ast} - c.c.)=2Nz \sin{k}$ : 
$<z> = 0$ implies that the wave is
completely reflected,
and $<z(\tau)> > 0$ (or $<z(\tau)> < 0$) that the wave is only partially
reflected by the impurity. The latter regime is given by a complete 
rotation of the pendulum about its center, and can be considered as a
self-trapping of the angular momentum. Equivalently, there
is an effective energy barrier which forbids
the complete reflection of the incident wave, and preserve its coherence.
The observation of a persistent current is associated to a superfluid 
regime of the DNLS equation.

The situation changes in the (quasi-)continuum limit (which
should not be confused with the $N \to \infty$ limit).
In this case phonons 
can be emitted only with quasi-momentum close to $k$, a condition which allows
the applications of the Landau superfluidity criteria. 
The crossover between continuum and
discrete limits is a very interesting one: 
We emphasize that the two-mode
dynamics is crucially related to
discreteness and nonlinearity, and disappears in the CNLS limit.
Yet, there is a striking analogy: in both cases (in the Landau and in the
"pendulum" criterion) the phonon emission out of the incident
wave (which, therefore, dissipates its energy)
can be inhibited by an effective energy barrier. The key difference
lies on the corresponding spectrum of the emitted phonons, which leads
to a completely different dynamics.

From the effective Hamiltonian
(\ref{ham}) we find a critical value $\Lambda_c$ 
for the pendulum oscillations
about its center given by
\begin{equation}
H(\phi(0),z(0))=2\epsilon/N:
\label{lambda-critico}
\end{equation}
when $\Lambda<\Lambda_c$, $z$ oscillates around $0$.
When $\Lambda=\Lambda_c$, asymptotically $z(\tau) \to 0$
and with $\Lambda>\Lambda_c$,
$<z(\tau)> \neq 0$. In Fig.1 we plot the average value of the
normalized angular momentum $L(\tau)/L_0$
for different values of $\Lambda/\Lambda_c$ and $z(0)=1$,
$\phi(0)=0$. The numerical solutions
of Eq.(\ref{DNLS}) are in agreement with 
the two-mode approximation 
(\ref{z-phi}), dashed line. 
In the inset of Fig.1 we plot the 
normalized angular momentum vs. time for different 
$\Lambda/\Lambda_c$.

The fixed points of Eqs.(\ref{z-phi}) can be found 
by solving for $\dot{z}=0$,
$\dot{\phi}=0$. In particular, we have the non-trivial
stationary (solitonic) solutions $\phi=(2m+1)\pi$,
$z=\pm \sqrt{1-\frac{2 \epsilon}{N \Lambda}}$ (with $m$ integer).
In Figs.2(a)-2(d) we compare the numerical and pendulum solutions for
the normalized angular momentum and 
the phase, in the cases of $\pi$ oscillations (a-b) 
and $\pi$ stationary points (c-d).

The previous discussion can be extended 
to the case of many impurities: replacing the ansatz 
Eq.(\ref{ansatz}) in Eq.(\ref{DNLS})
we obtain: $H = \frac{\Lambda z^2}{2} - \frac{2}{N}
\sqrt{1-z^2} \sum_n \epsilon_n\cos{(\phi+2 k n)}$.
After some algebra, the effective Hamiltonian becomes:
\begin{equation}
H = \frac{\Lambda z^2}{2} - \frac{2 \bar{\epsilon}}{N}
\sqrt{1-z^2} \cos({\phi+\alpha})
\label{ham-more-imp}
\end{equation}
with $\bar{\epsilon}$ and $\alpha$ given by the Fourier
transform of the defects distribution:
\begin{equation}
\label{conditions}
\bar{\epsilon}~ e^{i \alpha} = \sum_n \epsilon_n e^{2 i k n}.
\end{equation}
The critical value $\Lambda_c$ is given 
by Eq.(\ref{lambda-critico}) with the replacement 
$\epsilon \to \bar{\epsilon};~\phi_0 \to \phi_0 + \alpha$ \cite{nota1}.
It is also clear, from Eq.(\ref{conditions}), that the system becomes
transparent for some particular distribution of defects. 
For instance, with
an extended, step-like barrier
($\epsilon_n=constant$ for $\bar{n}_1 \le n \le \bar{n}_2$)
of length $10$ sites and with 
$2k=\pi/5$, we have $\bar{\epsilon}=0$. 
 
All the predicted regimes discussed so far have been 
found to be in agreement with our full numerical analysis. 
We considered different uniform random distributions of defects 
(e.g. all the $\epsilon_n$ positive or negative 
or with zero mean value). The critical values of the nonlinearity 
found from the numerical solution of Eq.(\ref{DNLS}) and the 
comparison with the theoretical prediction for $\Lambda_c$ from 
Eq.(\ref{lambda-critico}) is shown in Fig.3. 
In the inset of Fig.3 we plot $L(\tau)/L_0$ 
as a function of time for various
$\Lambda$ and a random distribution of defects $\epsilon_n$. 
The excellent agreement between numerics and the 
solution of Eq.(\ref{ham-more-imp}),
and the robustness of the two-mode ansatz in the presence of 
an arbitrary distribution of defects,
opens to the possibility
of studying the competition between disorder (and Anderson localization) 
and nonlinearity  
from a new perspective. Eq.(\ref{ham-more-imp}) is 
analytically solvable,
yet still it contains all the essential 
ingredients to investigate the details of the superfluid - normal 
transition in the DNLS 
with impurities.

To conclude we briefly discuss
the limits for recovering the CNLS equation (in an annular geometry)
from the DNLS Eq.(\ref{DNLS}). 
Writing $\Lambda = 2 m g_0/\hbar^2 N$,
$\epsilon_n = V_n m L^2/\hbar N^2$ and
$t = m L^2/\hbar N^2 \tau$, with $V_n \equiv V(x=x_n)$ the defect potential
in $x_n$, $L$ the length of the annulus and
$\tau$ the dimensionless time entering
in Eq.(\ref{DNLS}), the CNLS is obtained in the limit $N \to \infty$.
In particular, the critical value for the pendulum oscillations
Eq.(\ref{lambda-critico}) becomes $\Lambda_c = V_n m L^2/\hbar N^3 \to 0$.
Therefore, approaching the continuous limit, the DNLS
pendulum regime collapses to a (strongly) self-trapped
state. This prevents the emission of phonons with opposite momenta with
respect to the incident wave, whose energy will be eventually dissipated
on a much longer time scale, according to the Landau argument.

Discussions with K. \O. Rasmussen are acknowledged.
This work has been supported by
the U.S. DOE and the Cofinanziamento MURST.

\begin{figure}[h]
\centerline{\psfig{
figure=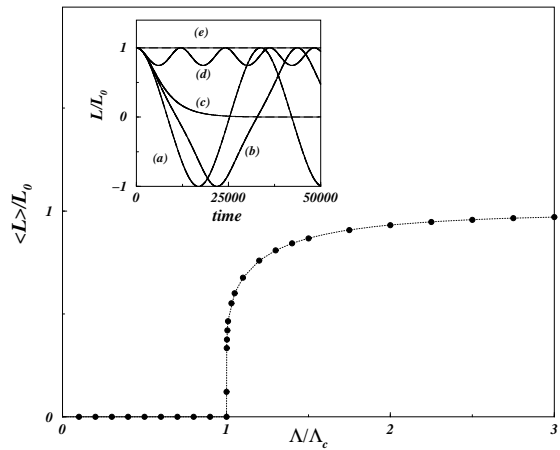,width=60mm,angle=270}}
\caption{
Average value of the angular momentum 
$L(\tau)$ (normalized to the 
initial value $L_0$) vs. the nonlinear coefficient $\Lambda/\Lambda_c$, 
($\Lambda_c=4 \epsilon / N$), with $\epsilon=0.01$, $N=100$, $z(0)=1$. 
The filled circles are the numerical solutions of Eq.(\ref{DNLS}), 
the dashed line is obtained from equations (\ref{z-phi}). 
Inset: normalized angular momentum vs. time for different values 
of $\Lambda/\Lambda_c=0.5,0.75,1,1.5,25$, 
respectively, corresponding to $(a),\cdots,(e)$. 
$\phi(0)=0$.}
\end{figure}

\begin{figure}[h]
\centerline{\psfig{
figure=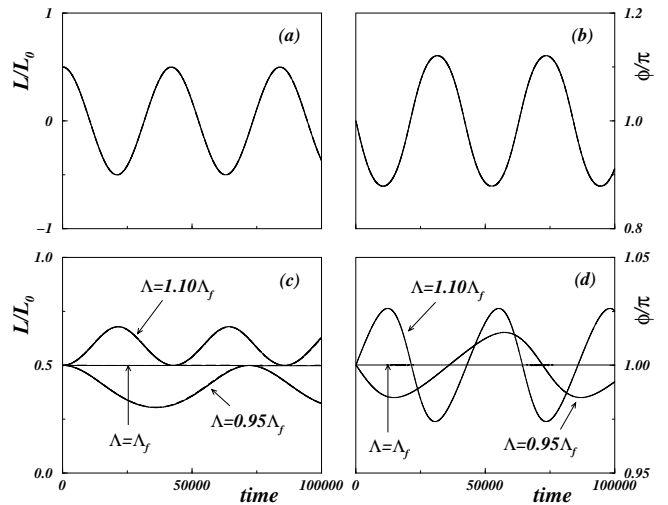,width=70mm,angle=270}}
\caption{Normalized angular momentum $(a)$ and phase $(b)$ vs. time for a 
$\pi$-state ($\Lambda/\Lambda_c=0.5$). 
In $(c)$ and $(d)$ we plot the same quantities for 
$\Lambda_f=(2\epsilon/N)/\sqrt{1-z^2(0)}$ (the stationary solution 
of Eq.(\ref{z-phi}) with $\phi=\pi$) and for 
$\Lambda/\Lambda_f=0.95,1.10$. In all cases solid (dashed) 
lines are for the numerical (variational) solution of Eq.(\ref{DNLS}) with
$\epsilon=0.01$, $N=100$, $z(0)=0.5$, $\phi(0)=\pi$.}
\end{figure}


\begin{figure}[h]
\centerline{\psfig{
figure=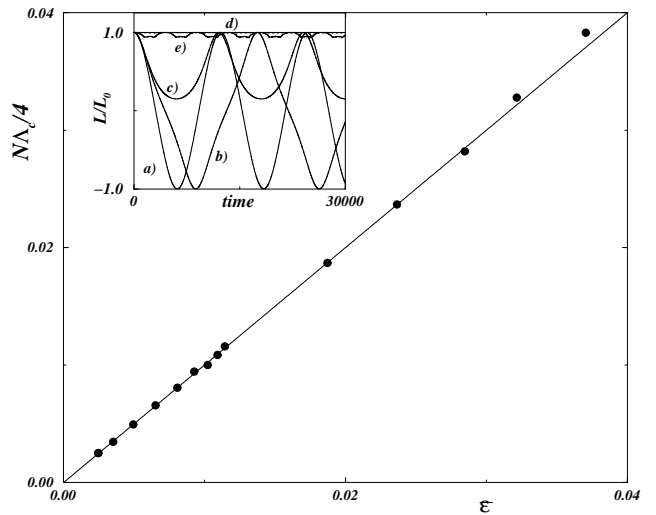,width=70mm,angle=270}}
\caption{Critical value $\Lambda_c$ vs. $\bar{\epsilon}$ 
for different random distributions of defects (and initial values 
$z(0)=1$, $\phi(0)=0$). 
Black circles: numerical solution of Eq.(\ref{DNLS}). 
Solid line: the analytical prediction 
$\Lambda_c=4\bar{\epsilon}/N$ with 
$\bar{\epsilon}$ given by Eq.(\ref{conditions}).  
Inset: 
angular momentum vs. time with a random 
distribution of defects for different values 
of $\Lambda/\Lambda_c=0.45,0.90,1.01,10,1000$ 
(corresponding to $(a),\cdots,(e)$). 
The sum of the strengths of the
random impurities is $0.1$. 
}
\end{figure}

\end{multicols}


\begin{thebibliography}{10}
\bibitem{nota1}
As an example, we consider the case of a gaussian
barrier $\epsilon_n=(\epsilon/\sqrt{\pi}\sigma)
e^{-(n-\bar{n})^2/\sigma^2}$ with width $\sigma$. The
effective Hamiltonian is 
$H = \frac{\Lambda z^2}{2} - (2 \epsilon \, e^{-k^2 \sigma^2}/N)
\sqrt{1-z^2} \cos({\phi+2k\bar{n}})$
with the critical value obtained from (\ref{lambda-critico}) now given by
$ \Lambda_c = 4 \epsilon e^{-k^2 \sigma^2} / N $.
\end{thebibliography}
\end{document}